\documentclass[runningheads]{llncs}
\usepackage{graphicx}
\usepackage{amsmath}
\usepackage{amsfonts}
\usepackage{floatrow}
\usepackage{hyperref}
\usepackage{arydshln}
\usepackage{svg}
\newfloatcommand{capbtabbox}{table}[][\FBwidth]

\def\usepng{1}  
\begin{document}
\title{Synthesising Rare Cataract Surgery Samples with Guided Diffusion Models}
\titlerunning{Synth. Cataract Samples with Guided Diffusion}
%
\author{
    Yannik Frisch\inst{1}\thanks{Corresponding author: yannik.frisch@gris.tu-darmstadt.de} 
    \and Moritz Fuchs\inst{1} 
    \and Antoine Sanner\inst{1}
    \and Felix Anton Ucar\inst{2} 
    \and Marius Frenzel\inst{2}
    \and Joana Wasielica-Poslednik\inst{2}
    \and Adrian Gericke\inst{2}
    \and Felix Mathias Wagner\inst{2}
    \and Thomas Dratsch\inst{3} 
    \and Anirban Mukhopadhyay\inst{1}
}
%

\authorrunning{
    Y. Frisch et al.
}
%

\institute{
    Technical University Darmstadt, Darmstadt, Germany \and
    Universitätsmedizin der Johannes Gutenberg-Universität Mainz, Mainz, Germany \and
    Uniklinik Köln, Cologne, Germany
}
\maketitle              
\begin{abstract}
Cataract surgery is a frequently performed procedure that demands automation and advanced assistance systems. However, gathering and annotating data for training such systems is resource intensive. The publicly available data also comprises severe imbalances inherent to the surgical process. Motivated by this, we analyse cataract surgery video data for the worst-performing phases of a pre-trained downstream tool classifier. The analysis demonstrates that imbalances deteriorate the classifier's performance on underrepresented cases. To address this challenge, we utilise a conditional generative model based on Denoising Diffusion Implicit Models (DDIM) and Classifier-Free Guidance (CFG). Our model can synthesise diverse, high-quality examples based on complex multi-class multi-label conditions, such as surgical phases and combinations of surgical tools. We affirm that the synthesised samples display tools that the classifier recognises. These samples are hard to differentiate from real images, even for clinical experts with more than five years of experience. Further, our synthetically extended data can improve the data sparsity problem for the downstream task of tool classification. The evaluations demonstrate that the model can generate valuable unseen examples, allowing the tool classifier to improve by up to 10\% for rare cases. Overall, our approach can facilitate the development of automated assistance systems for cataract surgery by providing a reliable source of realistic synthetic data, which we make available for everyone. 

\end{abstract}

\section{Introduction}

Cataract surgeries are amongst the most frequently performed treatments, with 4,000 to 10,000 annual operations per million people~\cite{wang2016cataract}. The high demand naturally asks for automation and advanced assistance systems and has seen increasing attention within the CAI community in recent years~\cite{al2019cataracts,grammatikopoulou2021cadis,roychowdhury2017identification}.

\begin{figure}
    \centering
    \if\usepng1
        \includegraphics[width=\textwidth]{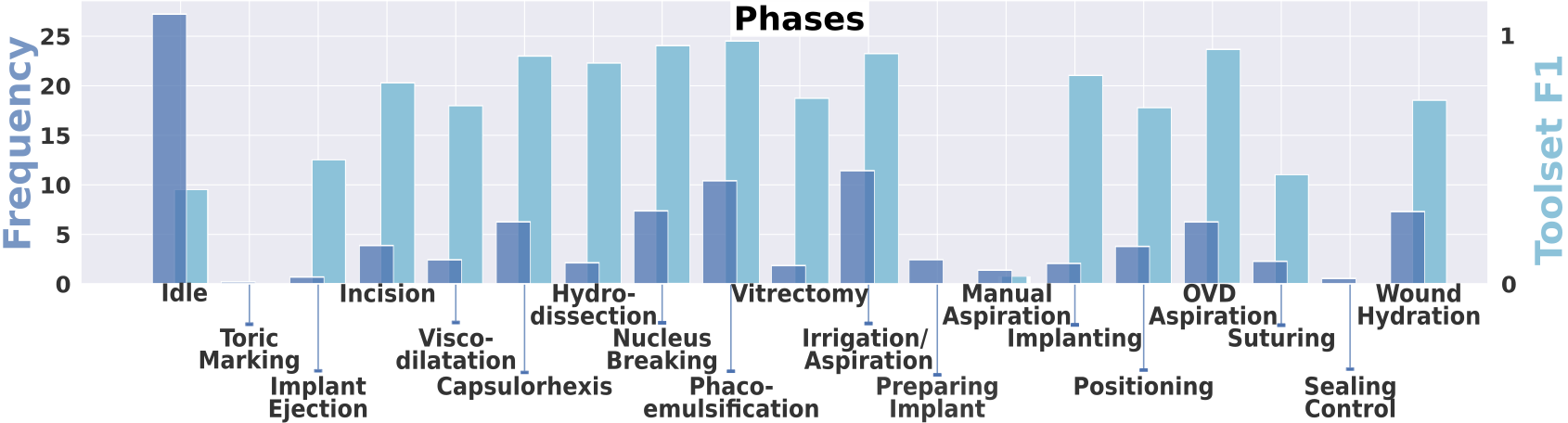}
    \else
        \includegraphics[width=\textwidth]{images/phase_dist_v2.eps}
    \fi
    \caption{Distribution of CATARACTS phases. Except for \textit{Idle} - which can appear anytime - all phases are displayed in the usual chronological order. The dataset yields severe class imbalances regarding the available frames per phase (darker blue), which results in performance drops for underrepresented phases (lighter blue).}
    \label{fig:cataracts_phases}
\end{figure}

Nevertheless, the publicly available data for training such systems is limited: given the nature of the surgeries, certain surgical \textit{phases} take more time than others. Further, there are variances in their length based on the surgeon's skill and the patient's particular needs. Since surgical \textit{tool} usage is strongly coupled with the surgical phase, certain phases and tools are shown more frequently than others, constituting an inherent imbalance in the data. As displayed in Figure~\ref{fig:cataracts_phases} for the CATARACTS dataset~\cite{al2019cataracts}, such imbalances impact the performance on downstream tasks, e.g. surgical phase prediction or tool classification. 

One cannot simply gather new data showing unusual tools to perform the required actions during a surgical step. Therefore, we must find different ways to represent them in the data and counteract the imbalance. The usual countermeasures in the form of oversampling can increase prediction accuracy~\cite{al2019cataracts,grammatikopoulou2021cadis,roychowdhury2017identification}. Still, they only alter the number of times an underrepresented sample is seen during training, resulting in a fragile representation of the tools and phases and hindering generalisation. Generative models~\cite{kalia2021co,pfeiffer2019generating,sommersperger2022surgical,uzunova2022systematic} can potentially solve this by synthesising unseen examples for underrepresented tool and phase combinations. 

Regarding image quality, generative models based on \textit{diffusion models} reached superior performance over alternative methods in the recent past~\cite{dhariwal2021diffusion,ho2020denoising,nichol2021improved,rombach2022high,song2020denoising}. 
Despite the successes, these models have not yet found application in Surgical Data Science. In the broader medical domain, they have been utilised to generate thorax CT scans~\cite{khader2022medical}, brain MRI~\cite{dorjsembe2022three,khader2022medical} and breast and knee MRI scans~\cite{khader2022medical}. 

Although these applications have shown promising results, there is a demand for conditional image generation for Surgical Data Science. Since most downstream applications consist of supervised methods, they require training targets, and the likelihood of unconditionally generating diverse samples for unusual cases is very low. For conditional generation with denoising diffusion models, \textit{classifier guidance} has been introduced recently~\cite{dhariwal2021diffusion,nichol2021glide} but requires computationally extensive parallel training of a separate classifier model. Instead, \textit{Classifier-Free Guidance} (CFG)~\cite{ho2022classifier} yields a simple trick to achieve class-constrained generative results with diffusion models. Conditional diffusion models have been applied to generate medical images from a few binary label inputs~\cite{muller2022diffusion,pinaya2022brain}. Peng et al.~\cite{peng2022generating} synthesise 3D volumes from 2D reference slides. Sagers et al.~\cite{sagers2022improving} have built on DALL$\cdot$E2 for the targeted generation of images of skin diseases, and Moghadam et al.~\cite{moghadam2023morphology} have generated histopathology images with genotype guidance.

\begin{figure}[t]
    \centering
    \if\usepng1
        \includegraphics[width=\textwidth]{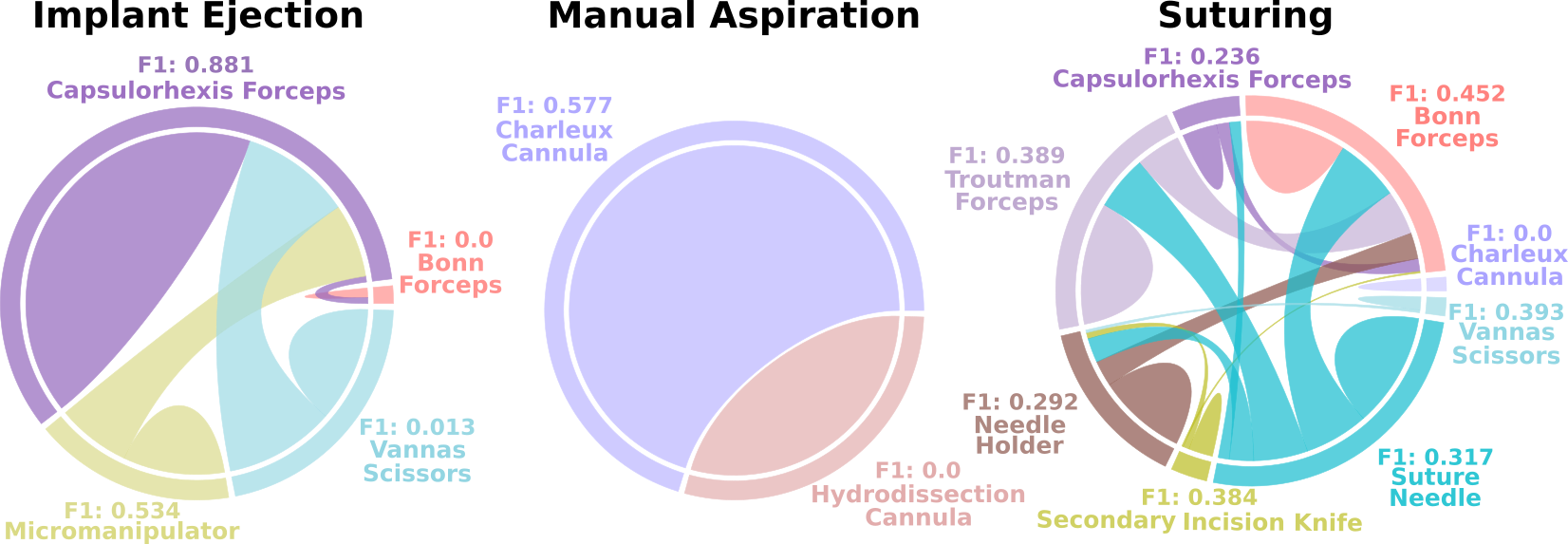}
    \else
        \includesvg[width=\textwidth]{images/worst_phases_v2.svg}
    \fi
    \caption{Ground truth toolset occurrence for the worst performing CATARACTS phases. Some toolsets, e.g. \textit{(Bonn Forceps, Capsulorhexis Forceps)} during \textit{Implant Ejection} (left), are rarely present and poorly detected, deteriorating the overall performance. We focus on such rare toolsets to generate new samples for a phase. E.g. for \textit{Manual Aspiration} (middle), we mainly want additional samples showing the \textit{Hydrodissection Canulla}. The chord diagram for the \textit{Suturing} phase (right) shows the complexity of such occurrences.}
    \label{fig:worst_phases}
\end{figure}

Precise conditioning beyond a few binary labels is crucial for synthesising valuable surgical data. Instead, we need to train a model that can generate diverse examples based on multi-class or multi-label conditions, e.g. certain surgical phases, combinations of surgical tools, or both. We show that using an adapted denoising diffusion model together with CFG can yield high-quality samples of cataract surgery data, even for rare cases such as the CATARACTS phase and tool combinations shown in Figure~\ref{fig:worst_phases}. 
    
To the best of our knowledge, ours is the first work combining CFG with diffusion models to efficiently \textbf{generate realistic cataract surgery data} with a complex underlying label structure. Additionally, we examine the cataract video data for the worst-performing phases of a pre-trained tool usage classifier. We then leverage the conditional denoising diffusion model to generate unseen samples for these phases. Our conditioned tools are \textbf{recognisable by the tool classifier} and are \textbf{hard to differentiate from real images}, even for clinicians with more than five years of experience. Further, we demonstrate how our synthetically extended data can \textbf{alleviate the data sparsity problem for the downstream task}. Overall, our evaluations show that the model can generate valuable examples to build the bridge to clinical application.

\section{Method}
The following section describes how we build our generative diffusion model and integrate CFG to generate cataract surgery frames conditioned on surgical phases and tools. Furthermore, we provide an analysis of the worst-performing surgical steps for a pre-trained tool classifier model. Finally, we demonstrate the sampling procedure using the generative model to improve the classifier.

\subsection{Denoising Diffusion Probabilistic Models}
The fundamental underlying idea of \textit{Denoising Diffusion Probabilistic Models} (DDPMs)~\cite{ho2020denoising} is a \textit{forward diffusion process} that gradually adds Gaussian noise to an image $x$. This process is defined by
$q(x_t|x_{t-1}) = \mathcal{N}(x_t; \sqrt{1 - \beta_t}x_{t-1}, \beta_t \mathbf{I})$,
which uses a pre-defined variance schedule $\{\beta_t \in (0, 1)\}^T_{t=1}$. Eventually, when $T\rightarrow\infty$, $x_T$ becomes equivalent to an isotropic Gaussian distribution. When we can learn the \textit{reverse process} $q(x_{t-1}|x_t)$, we can generate samples starting from a simple Gaussian noise $x_T \sim \mathcal{N}(\mathbf{0}, \mathbf{I})$. To achieve this, one can approximate the conditional probabilities by
$p_\theta(x_{t-1}|x_t) = \mathcal{N}(x_{t-1};\mu_\theta(x_t, t), \Sigma_\theta(x_t, t))$.
In practice and after some mathematical simplifications, this reduces the reverse process to estimating the noise $\epsilon_t$ between $x_t$ and $x_{t+1}$, as shown by Ho et al.~\cite{ho2020denoising}. Usually, the noise is parameterised by a UNet-type architecture $\epsilon_\theta$, optimised by minimising the simplified objective
\begin{equation}
    \label{eq:simple_objective}
    \begin{split}
        \mathcal{L}_t   &= \mathbb{E}_{t\sim[1,T],x_0,\epsilon_t}[||\epsilon_t - \epsilon_\theta(x_t, t)||^2] \\
                        &= \mathbb{E}_{t\sim[1,T],x_0,\epsilon_t}[||\epsilon_t - \epsilon_\theta(\sqrt{\alpha_t}x_0 + \sqrt{1 - \alpha}_t\epsilon_t, t)||^2]
    \end{split}
\end{equation}
where $\alpha_{1:T}\in (0,1]^T, \alpha_t = (1-\beta_t)\alpha_{t-1}$.

\textit{Denoising Diffusion Implicit Models} (DDIMs)~\cite{song2020denoising} are a generalisation of these formulations for non-Markovian, more efficient sampling. Instead of the complete Markov chain, they are defined on a reduced set of intermediate latents $\{x_{\tau_1},...,x_{\tau_S}\}$, where $[\tau_1,...,\tau_S]\subseteq[1,...,T]$. This reduction results in significantly fewer inference steps required to generate samples. 

\begin{figure}
    \centering
    \if\usepng1
        \includegraphics[width=\textwidth]{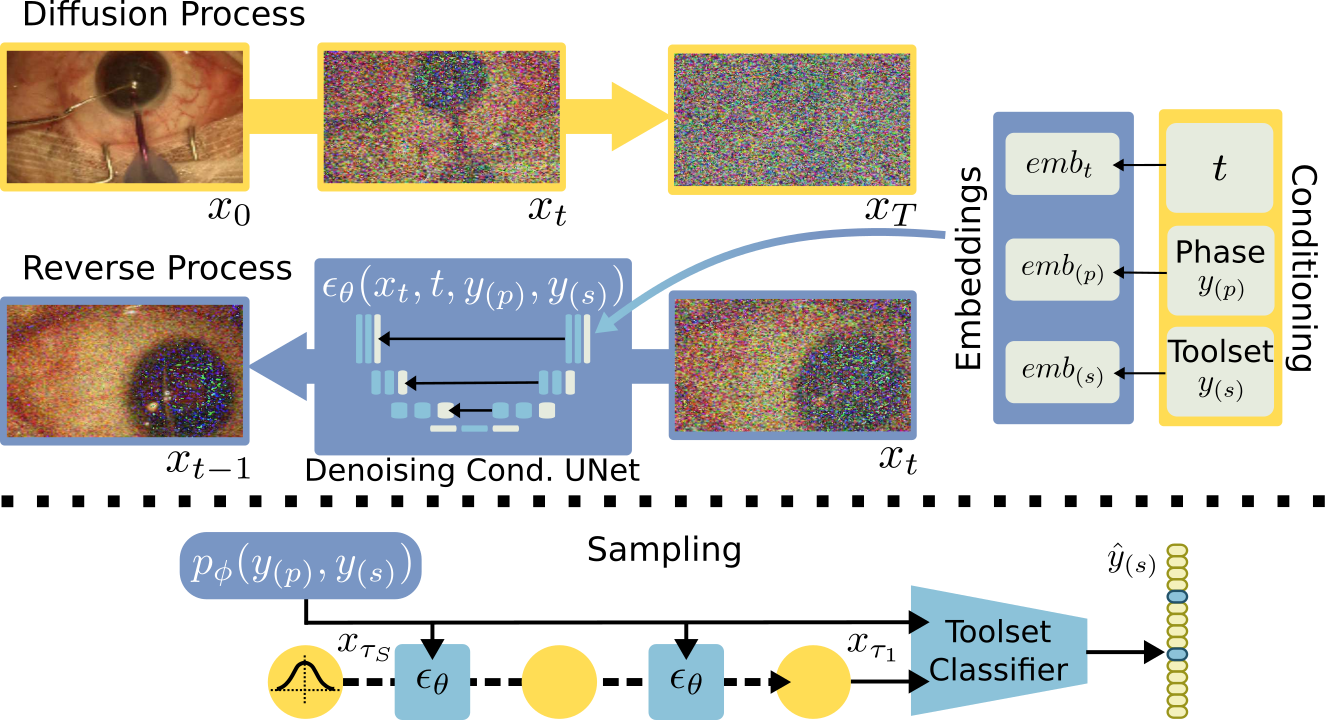}
    \else
        \includegraphics[width=\textwidth]{images/diffusion_v3.eps}
    \fi
    \caption{Illustration of the Conditional Denoising UNet. The model $\epsilon_\theta$ is trained to reverse the Diffusion Process, mapping a noisy sample $x_t$ to a less noisy $x_{t-1}$. Condition embeddings based on the surgical phase $y_{(p)}$ and the toolset $y_{(s)}$ are concatenated with the diffusion time step embedding and fed into every level of the UNet to guide targeted sample generation. We utilise the final model to synthesise realistic examples and improve the predictions $\hat{y}_{(s)}$ of a toolset classifier.}
    \label{fig:denoising_unet}
\end{figure}

\subsection{Classifier-Free Guidance}
By utilising \textit{Classifier-Free Guidance} (CFG)~\cite{ho2022classifier}, we can learn the unconditional model $p_\theta(x)$ and the model $p_\theta(x|y_{(p)}, y_{(s)})$ conditioned on phase $y_{(p)}$ and toolset $y_{(s)}$ using a \textit{single} neural network. The corresponding gradient is given by
\begin{equation}
    \begin{split}
        \nabla_{x_t} \log p(y_{(p)}, y_{(t)}|x_t) &= \nabla_{x_t} \log p(x_t|y_{(p)}, y_{(s)}) - \nabla_{x_t} \log p(x_t) \\
                                   &= - \frac{1}{\sqrt{1 - \bar{\alpha}_t}}(\epsilon_\theta(x_t, t, y_{(p)}, y_{(s)}) - \epsilon_\theta(x_t, t))
    \end{split}     
\end{equation}
This gradient yields
\begin{equation}
    \begin{split}
        \bar{\epsilon}_\theta(x_t, t, y_{(p)}, y_{(s)}) &= \epsilon_\theta(x_t, t, y_{(p)}, y_{(s)}) - \sqrt{1 - \bar{\alpha}_t} w \nabla_{x_t} \log p(y_{(p)}, y_{(s)}|x_t) \\
                                         &= \epsilon_\theta(x_t, t, y_{(p)}, y_{(s)}) + w (\epsilon_\theta(x_t, t, y_{(p)}, y_{(s)}) - \epsilon_\theta(x_t, t)) \\
                                         &= (w + 1)\epsilon_\theta(x_t, t, y_{(p)}, y_{(s)}) - w \epsilon_\theta(x_t, t)
    \end{split}
\end{equation}
where $w$ is a weighting hyperparameter. The weighted noise $\bar{\epsilon}_\theta(x_t, t, y_{(p)}, y_{(s)})$ can simply replace $\epsilon$ in Equation~\ref{eq:simple_objective}. We add an embedding module $emb_{(p)}$ to the UNet architecture, which converts categorical phase labels $y_{(p)}$ into one-hot encoded vectors to include them into the input. To simultaneously include tool labels in the form of (non-exclusive) binary vectors, we compute projections $emb_{(s)}$ of the same size as the time-step and phase label embeddings using stacked dense layers. All embeddings are concatenated as $\{emb_t(t), emb_{(p)}(y_{(p}), emb_{(s)}(y_{(s)})\}$ and fed to the conditional denoising UNet model together with the noisy image $x_t$. Figure~\ref{fig:denoising_unet} visualises the forward process, reverse process, and sampling procedure.

\subsection{Tool Usage Analysis \& Sample Generation}
\label{sec:phase_pred_sample_gen}
Following Roychowdhury et al.~\cite{al2019cataracts,roychowdhury2017identification}, we deploy a ResNet50 architecture to predict tools present in a given frame. An inspection of the phase-wise performance reveals that the model underperforms for underrepresented phases, as shown in Figure~\ref{fig:worst_phases}. Certain tool combinations are sparsely used in these phases, causing a significant drop in prediction performance. Appendix Figure~\ref{fig:tools_per_phase} displays the distribution of toolset labels for all surgical steps. We synthesise new examples for every phase to smooth out the distribution. We then re-train the classifier model on the original and extended data combined. Adding samples based on toolsets $y_s$ and phase labels $y_p$ requires a throughout pre-selection of query inputs due to the complex underlying latent structure. To automatise this process, we compute the joint probabilities $p_\phi(y_s, y_p)$ from the available CATARACTS annotations. We can then generate rare cases for a given phase by sampling tool labels from the inverse of $p_\phi(y_s,y_p)=p(y_p)p(y_s|y_p)$. This ensures we synthesise examples of underrepresented cases.

\section{Experiments \& Results}
In this section we explain our experimental setup, demonstrate the synthesis of high-quality samples and show how these can improve the downstream model's performance on challenging phases.

\begin{figure}
    \centering
    \if\usepng1
        \includegraphics[width=\textwidth]{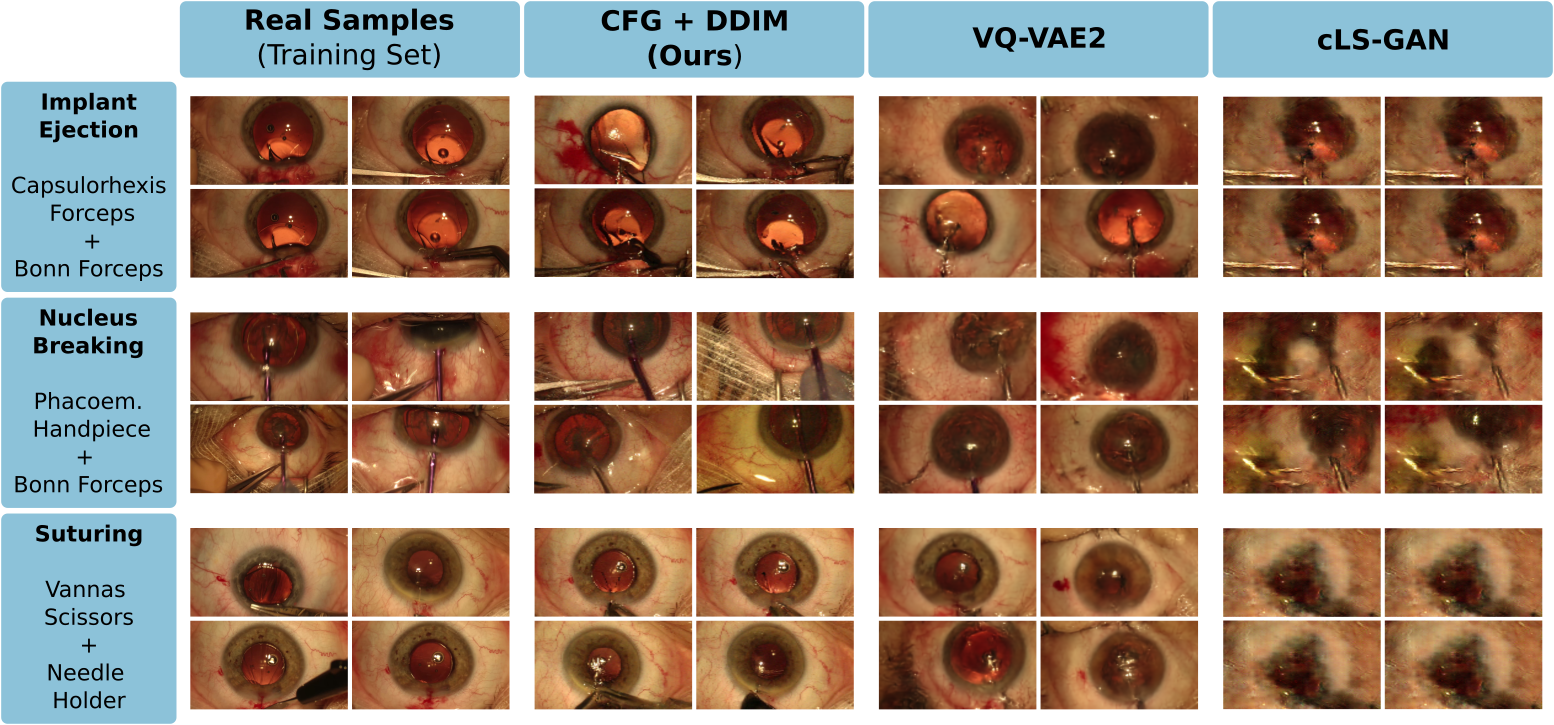}
    \else
        \includegraphics[width=\textwidth]{images/qual_examples.eps}
    \fi
    \caption{Qualitative examples for the least common toolsets of the three most challenging phases of CATARACTS. The proposed method produces superior results compared to the baselines, which struggle with the rare combinations of tools and phases shown. The realistically generated tools are especially noteworthy.}
    \label{fig:qual_samples}
\end{figure}

\subsection{Experimental Setup \& Dataset}
 To evaluate the quality of synthesised images, we generate 30,000 samples with phase and toolset conditions sampled from $p^{-1}_\phi(y_s, y_p) = (1 - p_\phi(y_s, y_p))/\sum(1 - p_\phi(y_s, y_p))$, as explained in Section \ref{sec:phase_pred_sample_gen}. The resulting number of examples is close to the test split size of CATARACTS sampled at 3 FPS. We compare the proposed approach to state-of-the-art baselines for conditional generative modelling: A conditional LS-GAN (cLS-GAN)~\cite{mao2017least,mirza2014conditional} and VQ-VAE2~\cite{razavi2019generating}. For the latter, we deploy a PixelSNAIL prior~\cite{chen2018pixelsnail} for bottom- and top-level features. Every model is trained on two NVIDIA A40 GPUs for 500 epochs with about 45,000 training examples each. Other hyperparameters vary for each model and can be accessed next to the code to reproduce our results and the generated data at \url{https://github.com/MECLabTUDA/CataSynth}. We take $\tau_S = 200$ denoising steps using the DDIM formulation~\cite{song2020denoising} to generate our images, yielding a reasonable inference speed and sample quality trade-off. The displayed and evaluated images are generated with a CFG weight of $\omega=2.0$. We use a random chance of $p=0.1$ for the unconditional model during training. All models are trained on images of $128 \times 128$ pixels and up-sampled with bilinear interpolation to $270 \times 480$ for displaying purposes.
 
 \subsection{Quantitative Image Quality}
We deploy a variety of quantitative metrics to assess the quality of generated images, for which Table ~\ref{tab:quantitatice_image_quality} lists the results. Firstly, \textbf{FID}\cite{binkowski2018demystifying} and \textbf{KID}\cite{binkowski2018demystifying} compare the spatial distribution of the synthesised images with the training set distribution of CATARACTS. Further, we use the classifier from Section~\ref{sec:phase_pred_sample_gen} to obtain the Inception Score (\textbf{IS}) of the generated images. By using the scores of a classifier trained for tool recognition, this metric yields a measurement of tool realism. Additionally, we evaluate the F1 score for the pre-trained classifier identifying tools in the conditionally generated images. We denote this metric as \textbf{CF1}. Lastly, we also compute the perceptual \textbf{LPIPS diversity}~\cite{zhang2018unreasonable} to catch mode collapses, a common problem with generated models, leading to reduced image variability.  In summary, our model generates images of superior quality regarding spatial properties, tool label preservation and diversity. 
 
\begin{table}
    \centering
    \begin{tabular}{|c||c|c|c|c|c|}
         \hline
         Method & FID ($\downarrow$) & KID ($\downarrow$) & CF1 ($\uparrow$) & IS ($\uparrow$) & LPIPS div. ($\uparrow$) \\
         \hline
         \hline
         cLS-GAN & 284.5 & 0.319 $\pm$ 0.005 & 0.000 & 1.376 $\pm$ 0.009 & 0.559 $\pm$ 0.119 \\
         \hline
         VQ-VAE2 & 88.9 & 0.096 $\pm$ 0.002 & 0.089 & 2.149 $\pm$ 0.035 & 0.502 $\pm$ 0.061 \\
         \hline
         CFG + DDIM & $\mathbf{43.7}$ & $\mathbf{0.030 \pm 0.002}$ & $\mathbf{0.433}$ & $\mathbf{6.428 \pm 0.115}$ & $\mathbf{0.595 \pm 0.070}$\\
         \hline
    \end{tabular}
    \caption{Quantitative image quality evaluation.}
    \label{tab:quantitatice_image_quality}
\end{table}

\subsection{Qualitative Results \& User Study}
Figure~\ref{fig:qual_samples} displays generated samples for the rarest toolsets of CATARACTS' three most challenging phases. Additional examples for the other phases and randomly chosen toolsets are displayed in the Appendix. The qualitative results reflect the quantitative metrics and illustrate our method's superior image quality and tool preservation. Besides qualitative evaluations, we conduct a user study to assess the realism of our generated images. Therefore, we let six clinicians survey 50 generated and 50 real images of size 235 $\times$ 132 pixels in a randomised side-by-side view. In every example, both images show the same phase and tool combination, and participants must distinguish the real image from the synthesised sample. We grouped the participants by domain experience, with domain experts (DE) having more than five years of domain expertise in cataract surgery. On average, the miss rate or false classification rate (FR) was 0.61. This result translates into \textbf{clinical experts favouring the generated images 61\% of the time} and highlights how realistic they appear. Their answers have an average Matthews Correlation Coefficient (MCC) of -0.216, showing low validity of the subjects' binary decisions. Individual MCC and FR scores are given in Table~\ref{tab:user_study}. 

\begin{table}
    \centering
    \begin{tabular}{|c||c|c|c|c|c|c|}
         \hline
         Clinician & NDE1 & NDE2 & NDE3 & DE1 & DE2 & DE3\\
         \hline
         \hline
         MCC & -0.961 & -0.288 & -0.201 & -0.233 & 0.098 & 0.288\\
         \hline
         FR & 49/50 & 32/50 & 30/50 & 31/50 & 23/50 & 18/50\\
         \hline
    \end{tabular}
    \caption{Results for user study on image realism.}
    \label{tab:user_study}
\end{table}

\subsection{Downstream Tool Classification}

\begin{figure}
    \begin{floatrow}
        \ffigbox[0.4\textwidth]{%
            \centering
            \if\usepng1
                \includegraphics[width=0.4\textwidth]{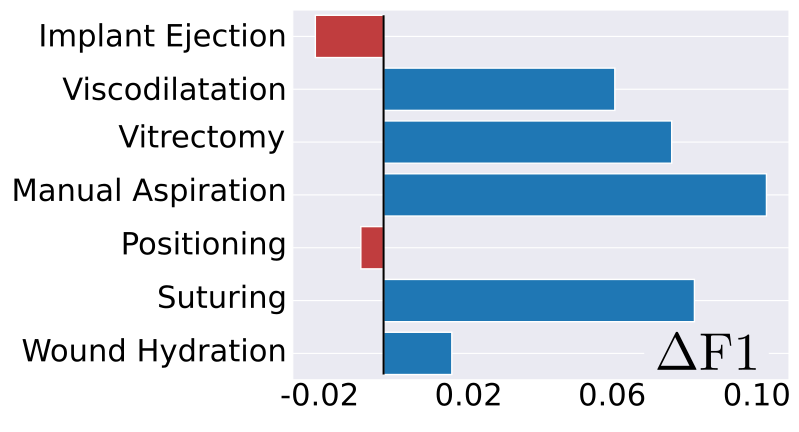}
            \else
                \includegraphics[width=0.4\textwidth]{images/performance_change.eps}
            \fi
        }{%
            \caption{Phase-wise performance changes for critical phases after re-training including synthetic data.}
            \label{fig:phase_pred}
        }
        \capbtabbox{%
            \begin{tabular}{cccc} \hline
                Data & F1 ($\uparrow$) & AUROC ($\uparrow$)& Acc. ($\uparrow$)\\ 
                \hline
                Original & 0.897 & 0.986 & 0.9921 \\
                Extended & $\mathbf{0.916}$ & $\mathbf{0.989}$ & $\mathbf{0.9924}$ \\
                \hdashline
                Synthetic (CAS) & 0.299 & 0.681 & 0.9502 \\
                \hline
            
            \end{tabular}
        }{%
            \caption{Tool-set prediction performance on the CATARACTS test split for different types of data.}%
            \label{tab:phase_pred}
        }
    \end{floatrow}
\end{figure}

Finally, we re-train the tool-set classifier on a combined dataset of the original and the synthesised samples. As shown in Table~\ref{tab:phase_pred}, re-training on the combined data (\textit{Extended}) improves the tool-set classifier's prediction performance on the original test data compared to training solely on the original training data (\textit{Original}). For completeness, we also report the performance from fitting the classifier exclusively on the synthetic data and evaluating it on the test split of CATARACTS, denoted \textit{CAS}~\cite{razavi2019generating}. Figure~\ref{fig:phase_pred} displays the individual differences in the classifier's F1 scores for the originally worst-performing phases. Extending the data with synthetic samples yields \textbf{performance gains for five of the seven most critical phases}.

\section{Conclusion}
We present a generative model based on denoising diffusion models and classifier-free guidance, powerful enough to synthesise cataract surgery images that are hard to distinguish for a pre-trained tool classifier and clinical experts. For underrepresented phases, state-of-the-art baselines tend to produce frames that show eyes without correct anatomy or barely recognisable tools, resulting in a significant performance gap. Distortions in the dataset further deteriorate their learning. On the contrary, we demonstrate that the proposed approach outperforms these baselines in terms of image quality and tool preservation. As a limitation of our approach, we found that the generalisation capabilities must be strengthened to generate unreasonable samples, e.g. completely wrong tools during a phase. Such samples can happen if they are present in the data. Though, a targeted generation would require a more substantial representation. Additionally, while tool realism is significantly better for the proposed method, the \textit{CF1} and \textit{CAS} scores indicate that it can be further improved.
Besides, the underlying class imbalances and lack of available data are even more severe for the downstream task of anatomy and tool segmentation. In future work, we will extend the proposed method to generate segmentation targets, temporally connected data and deploy a tighter structure for conditioning. Overall, we are the first to have shown how conditional diffusion models can successfully be applied to mend data sparsity and generate high-quality cataract surgery images suitable for clinical application. These improvements can bring computer-assisted cataract surgery one step closer to the next level of automation.

%
%
%
\bibliographystyle{splncs04}
\bibliography{refs}

\newpage
\appendix
\section{Synth. Cataract Samples with Guided Diffusion - Supplementary}
\subsection{Available Frames per Tool per Phase}
\label{sec:tool_dist}

\begin{figure}
    \centering
    \includegraphics[width=\textwidth]{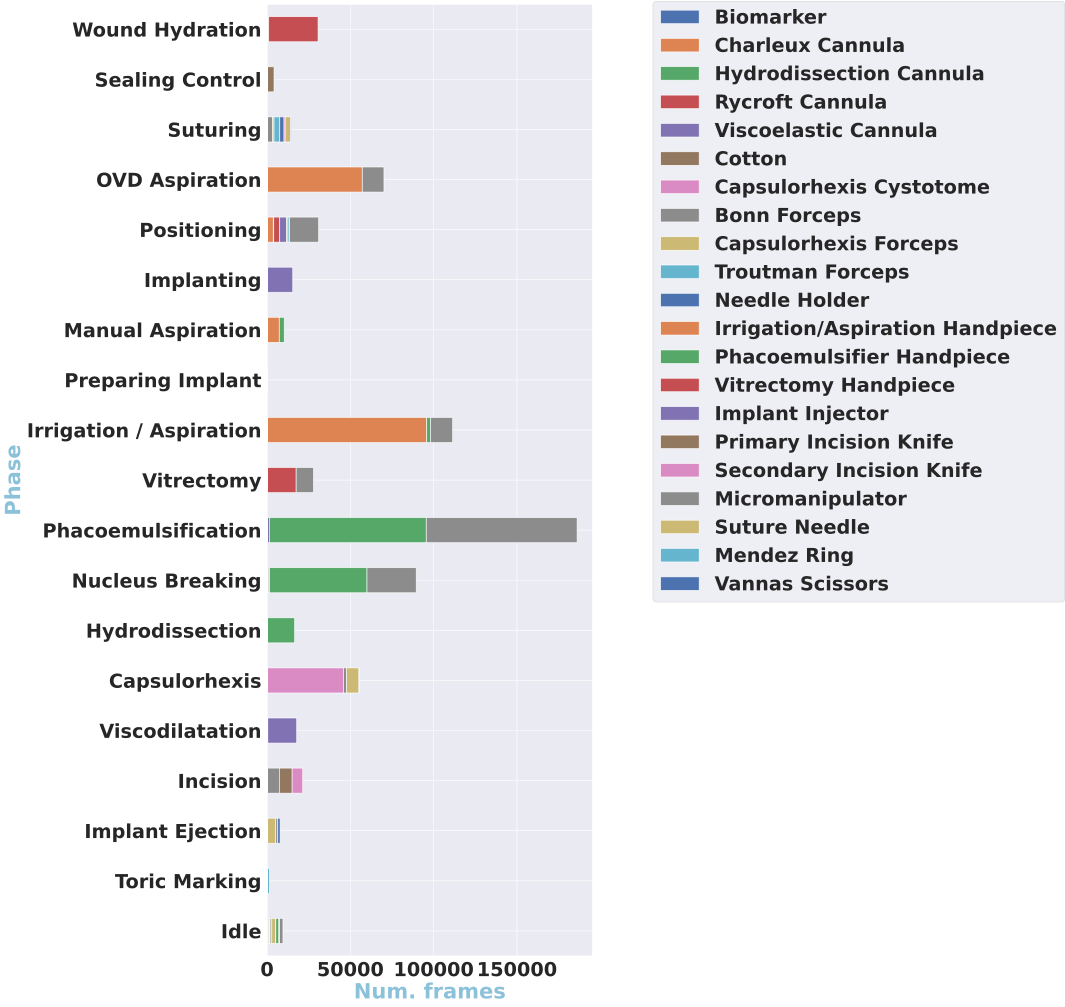}
    \caption{Imbalanced tool distribution for CATARACTS surgical phases.} 
    \label{fig:tools_per_phase}
\end{figure}

\newpage
\subsection{Further Qualitative Examples}
\label{sec:more_qual}

\begin{figure}
    \centering
    \if\usepng1
        \includegraphics[width=\textwidth]{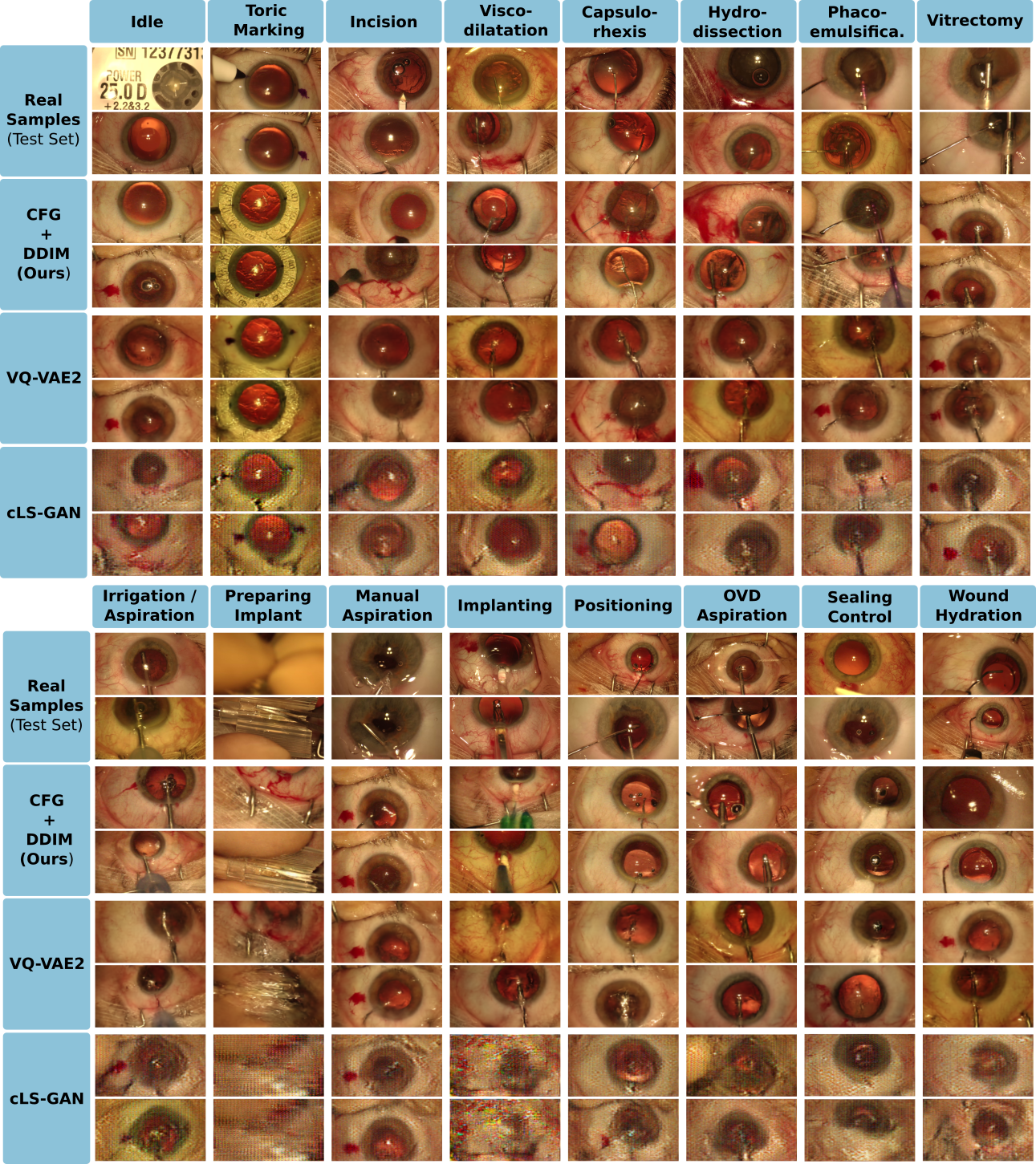}
    \else
        \includegraphics[width=\textwidth]{images/more_qual_examples_test_set.png}
    \fi
    \caption{Further qualitative examples.}
    \label{fig:downstream_model}
\end{figure}

\end{document}